\documentclass{INTERSPEECH2023}
\usepackage{multirow}
\usepackage{wasysym}
\usepackage{bbm}
\usepackage{url}
\usepackage{amssymb}
\usepackage{pifont}
\newcommand{\cmark}{\ding{51}}%
\newcommand{\xmark}{\ding{55}}%
\usepackage{spconf, amsmath,amsfonts,graphicx}
\usepackage{amssymb,textcomp,mathtools}
\usepackage{bm,upgreek,algorithm,hyperref}
\usepackage{multirow,booktabs,hhline,array}
\usepackage{cite,url,makecell,setspace}
\usepackage{xcolor}
\usepackage{cite}

\usepackage{lipsum}

\newcommand\blfootnote[1]{%
  \begingroup
  \renewcommand\thefootnote{}\footnote{#1}%
  \addtocounter{footnote}{-1}%
  \endgroup
}

\interspeechcameraready

\title{High Fidelity Speech Enhancement with Band-split RNN}
\name{Jianwei Yu$^*$,  Hangting Chen$^*$, Yi Luo, Rongzhi Gu, Chao Weng}
\address{
  Tencent AI Lab, Audio and Speech Signal Processing Oteam
  }
\email{\{tomasyu, oulyluo, erichtchen, lorrygu, cweng\}@tencent.com}

\begin{document}

\ninept
\maketitle

\begin{abstract}
Despite the rapid progress in speech enhancement (SE) research, improving the intelligibility and perceptual quality of desired speech in noisy environments with interfering speakers remains challenging.
This paper attempts to achieve high-fidelity full-band SE and personalized SE (PSE) by modifying the recently proposed band-split RNN (BSRNN) model.
To reduce the negative impact of unstable high-frequency components in full-band speech recording, we perform bi-directional and uni-directional band-level modeling to low-frequency and high-frequency subbands, respectively. 
For the PSE task, an additional speaker enrollment module is added to BSRNN to make use of the target speaker information for suppressing the interfering speech.
Moreover, we utilize a MetricGAN discriminator (MGD) and a multi-resolution spectrogram discriminator (MRSD) to further improve the human auditory perceptual quality of the enhanced speech. 
Experimental results show that our system outperforms various top-ranking SE systems, achieves state-of-the-art (SOTA) SE performance on the DNS-2020 test set, and ranks among the top 3 in the DNS-2023 challenge on the PSE task.


\end{abstract}
\begin{keywords}
Band-split RNN, Full-band speech enhancement, Personalized speech enhancement
\end{keywords}

\section{Introduction}
\label{sec:intro}
Speech enhancement (SE)\blfootnote{$^*$ Equal contribution} is an important task in speech communication that aims to improve the subjective and objective quality of speech signals. 
Such improvements can not only be helpful for human listeners to better understand the contents but also be beneficial for machine listeners to generate more accurate transcriptions. 
In recent few years, deep learning based SE models have made a great progress due to the emergence of novel model architectures \cite{dccrn, fullsubnet, BSRNN, demucs, li2022glance, wang2021tstnn}, efficient data simulation and model training pipelines \cite{xia2020weighted, fu2019metricgan}, and international challenges\cite{reddy2021icassp, dubey2023icassp} with comprehensive evaluation metrics and large-scale training data. 
Due to the rising demand for high-fidelity (Hi-Fi) speech in online conferencing systems and high-definition live streaming, recent developments in speech enhancement models have attempted to handle super wide-band (24 kHz) and full-band (48 kHz) speech signals\cite{S-DCCRN, zhang2022multi}.
Moreover, since conventional SE models cannot suppress the speech of interfering speakers, personalized SE (PSE) models have also been proposed \cite{wang2018voicefilter, giri2021personalized, ju2022tea, ju2023tea} to isolate the target speaker's voice from interfering speech using target speaker's enrollment speech.

Although previous SE and PSE models can largely improve the quality of the desired speech signals, the performance of these models on full-band signals is restricted by the following two aspects.
First, in real-time speech communication, different devices have various effective sample rates and frequency responses. 
Some devices can precisely capture relatively high-frequency ($>$8 kHz) information of the speech signal, while others may introduce significant distortion in the high-frequency range. 
Such unpredictable device-dependent distortion on high-frequency components can degrade the performance of full-band speech enhancement.
Second, the conventional training objectives of SE models, such as signal-to-noise ratio (SNR) and frequency-domain mean square error (MSE), are not closely related to human auditory perception and other objective perceptual metrics, which can lead to sub-optimal SE performance. 

\begin{figure*}[!htb]
	\centering
	\includegraphics[width=1\linewidth]{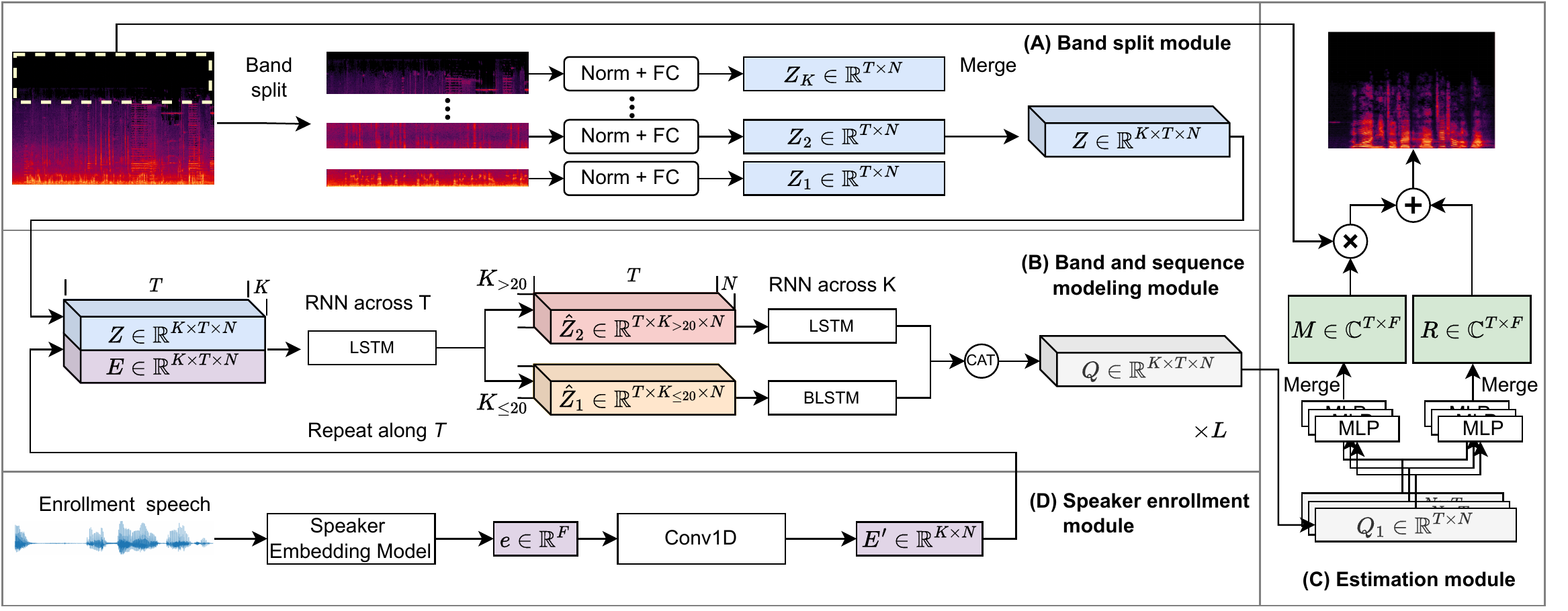}
    \caption{The diagram of the proposed full-band PSE system. (A) The band split module. (B) The sequence and band modeling module. (C) The estimation module. (D) The speaker enrollment module.}
\label{fig:BSRNN}
\end{figure*}

This paper aims to tackle the problems mentioned above by applying the \textit{band-split RNN (BSRNN)} model, which was recently introduced in \cite{BSRNN}, to full-band SE and PSE tasks.
In the modified BSRNN, the input signal's spectrogram is initially divided into a collection of frequency bands, which are then converted into subband features using band-specific fully connected (FC) layers.
For full-band SE and PSE tasks, we perform bi-directional band-level modeling for subbands below 8 kHz and uni-directional modeling for subbands above 8 kHz to mitigate the effect of high-frequency distortion introduced by the devices.
For PSE task, an additional speaker enrollment module is added to BSRNN to make use of the target speaker information.
In addition, we use a modified MetricGAN-based training objective \cite{fu2019metricgan, fu2021metricgan+} to directly optimize the model towards the widely-used perceptual evaluation of speech quality (PESQ) score.
Furthermore, to improve the human auditory perceptual quality of the enhanced speech, a multi-resolution spectrogram discriminator (MRSD) \cite{jang2021univnet} is also adopted in model training.
Experimental results show that our BSRNN system outperforms various top-ranking benchmark systems in SE and achieves SOTA results on the DNS-2020 non-blind test set in both offline and online scenarios.
For the PSE task, the proposed system achieves better PDNSMOS scores on the 4th DNS-2022 blind test set than the previous SOTA system and ranks among the top 3 in the DNS-2023 challenge on both speaker phone and headset tracks.

This paper is an extension of our two-page grand challenge report \cite{YUDNS2023} for ICASSP DNS 2023 Challenge. Both online and offline systems for SE and PSE tasks are considered in this paper, while the two-page grand challenge report only focuses on the online PSE task. More details of the model design, training objectives, and experimental results are added in this paper.

The paper is organized as follows: Section~\ref{sec:model} presents the proposed BSRNN architecture. The training objectives are discussed in Section~\ref{sec:obj}. Section~\ref{sec:config} introduces the data simulation and training configurations. Experimental results and analysis can be found in Section~\ref{sec:result}, and Section~\ref{sec:conclusion} offers the conclusion.

\section{Model Architecture}
\label{sec:model}
The proposed system shown in Figure~\ref{fig:BSRNN} is a frequency-domain model composed of a \textit{band split module}, a \textit{band and sequence modeling module}, a \textit{mask estimation module}, and an additional \textit{speaker enrollment module} for PSE purposes.
As shown in Figure~\ref{fig:BSRNN} (A), the band split module explicitly splits the complex-valued spectrogram of the noisy input $\vec{X}$ into a set of non-overlapped frequency bands $\{\vec{B}_i\}_{i=1}^{K}$ and generates a series of subband features $\{\vec{Z}_i\}_{i=1}^{K}$ with band-specific FC layers.
Residual long short-term memory networks (LSTMs) are utilized in the band and sequence modeling module to conduct interleaved modeling at both the band and sequence levels of the subband features. 

As discussed in Section~\ref{sec:intro}, different recording devices can introduce unpredictable distortions and information lost in high-frequency components (normally $>$8 kHz) due to different frequency responses. For example, the frequency components above 16 kHz are missing in the spectrogram shown in Figure~\ref{fig:BSRNN} (A). 
Hence, the bi-directional intra-band modeling used in original BSRNN model \cite{BSRNN} can be sub-optimal, as it can propagate high-frequency distortions to lower frequencies and thereby compromise the overall enhancement performance. To attenuate the effects of the high-frequency distortions in full-band enhancement, we split the subband features $\hat{\vec{Z}}$ to  $\hat{\vec{Z}_1}$ and $\hat{\vec{Z}_2}$, where $\hat{\vec{Z}_1}$ contains the subband features below 8 kHz and $\hat{\vec{Z}_2}$ contains subbands above 8 kHz. Then we perform bi-directional band-level modeling on  $\hat{\vec{Z}_1}$ and uni-directional modeling on $\hat{\vec{Z}_2}$ as follows: 
\begin{align}
    \vec{Q}_1, (\vec{h}, \vec{c}) = \text{BLSTM}(\hat{\vec{Z}}_1); \ \ \  \vec{Q}_2 = \text{LSTM}(\hat{\vec{Z}}_2, (\vec{h}, \vec{c}))
\end{align}
where $\vec{h}$ and $\vec{c}$ are the hidden and cell state of the low-to-high direction obtained from the band-level modeling of $\hat{\vec{Z}}_1$. In the rest of the paper, we will use \textit{BSRNN-S} to denote the BSRNN model with split band-level modeling. 

After band and sequence modeling, the estimation module uses band-specific MLPs to predict the complex-value T-F mask $\vec{M} \in \mathbb{C}^{F \times T}$. 
Regarding the artifacts brought by the complex masks, an MLP is additionally used to directly predict the target speech's complex-valued residual spectrogram $\vec{R}\in \mathbb{C}^{F \times T}$. The final enhanced speech is obtained by:
\begin{align}
    \bar{\vec{S}} =  \vec{M} \odot \vec{X} + \vec{R}
\end{align}
where $\odot$ denotes the complex-valued Hadamard product and $ \bar{\vec{S}}$ represents the enhanced spectrogram.

In PSE task, a speaker enrollment module is utilized to extract the target speaker information from the enrollment speech to eliminate interfering speech.
Specifically, the enrollment speech is a pre-recorded 5 to 10 seconds clean or noisy speech segment from the target speaker, which is used to compute the speaker embedding $\mathbf{e}\in\mathbb{R}^{F}$. 
In the proposed model, we adopt a pretrained speaker embedding model\cite{wang2023wespeaker} based on ResNet34 with frozen model parameters to compute the speaker embedding $\mathbf{e}$.
The speaker representation $\mathbf{E}\in\mathbb{R}^{K\times T \times N}$ is then obtained by transforming the speaker embedding $\mathbf{e}$ using band-specific 1-D convolutional layers with \textit{Tanh} activation function and repeated across the time axis. 
Then, we concatenate the subband feature $\mathbf{Z}$ and the speaker representation $\mathbf{E}$ along the feature dimension to apply band and sequence-level modeling.
Throughout the remainder of the paper, we will refer to the personalized BSRNN model as pBSRNN.

\vspace{-0.25cm}
\section{Training Objectives}
\label{sec:obj}
The training objective of our model contains three parts. 

\noindent\textbf{Multi-resolution Frequency Loss}:
The first part is a multi-resolution (MR) frequency-domain loss $\mathcal{L}^{\texttt{MR}}$ designed by combining the power-compressed amplitude mean average error (MAE) loss and the complex-valued MAE loss with short-time Fourier transform (STFT) windows ranging from 10 ms to 40 ms: 
\begin{align}
    \mathcal{L}^{\texttt{MR}} = \frac{1}{I}\sum_{i} \left( \Arrowvert|\vec{S}_i|^{p} - |\bar{\vec{S}}_{i}|^{p}\Arrowvert_1 + \Arrowvert\vec{S}_{i} - \bar{\vec{S}}_{i}\Arrowvert_1\right) 
\end{align}
where ${\bar{\vec{S}}}_{i}$ and ${\vec{S}}_{i}$ denote the complex spectra of the estimated and target speech signal, respectively,  $i$ is the index of the STFT window size from [10, 20, 30, 40] ms.  We use $p=0.3$ in our implementation.


\noindent\textbf{MetricGAN}: To directly improve the system performance on the PESQ score, adversarial training based on MetricGAN (MGD) \cite{fu2019metricgan, fu2021metricgan+} is adopted.
The generator of MGD is the BSRNN model and the discriminator in MGD $D(\cdot)$ attempts to predict the normalized PESQ score:
\begin{align}
    Q_{\text{PESQ}}(\bar{\vec{S}}, {\vec{S}}) = \frac{\text{PESQ}(\bar{\vec{S}}, \vec{S}) + 0.5}{5} 
\end{align}
Note that the wide-band PESQ score is used in all experiments. Two least-square GAN (LSGAN) \cite{mao2017least} style training objectives are then designed for the generator and discriminator:
\begin{align}
    \mathcal{L}_{d}^{\texttt{MGD}} &= (1 - D({\vec{S}}, \vec{S}))^{2} + (Q_{\text{PESQ}}(\bar{\vec{S}}, \vec{S}) - D(\bar{\vec{S}}, \vec{S}))^{2} \\ \nonumber 
        &+ (Q_{\text{PESQ}}({\vec{X}}, {\vec{S}} ) - D({\vec{X}}, {\vec{S}}))^{2} \\
    \mathcal{L}_{g}^{\texttt{MGD}} &= (1 - D(\bar{\vec{S}}, \vec{S}))^{2} + 0.5 \cdot \mathcal{L}^{\texttt{MR}}
\end{align}
where $\mathcal{L}_{g}^{\texttt{MGD}}, \mathcal{L}_{d}^{\texttt{MGD}}$ denote the generator loss and the discriminator loss, respectively. Note that the discriminator loss contains three signal pairs from noisy input, clean target and estimated target to stabilize the training process.

\noindent\textbf{Multi-resolution Spectrogram Discriminator}:
To further enhance the perceptual quality of the output speech, we also adopt the multi-resolution spectrogram discriminator (MRSD) \cite{jang2021univnet} in model training: 
\begin{align}
    &\mathcal{L}_{d}^{\texttt{MRSD}} = \frac{1}{K} \sum_{k} \left( (1 - D_{k}(\vec{S}))^{2} +D_{k}(\bar{\vec{S}})^{2} \right)  \\
    &\mathcal{L}_{g}^{\texttt{MRSD}} = \frac{1}{K} \sum_{k} (1 - D_{k}(\bar{\vec{S}}))^{2}  + \mathcal{L}^{\texttt{MR}} + \mathcal{L}^{\texttt{MMEL}}
\end{align}
where  $\mathcal{L}_{d}^{\texttt{MRSD}}, \mathcal{L}_{g}^{\texttt{MRSD}}$ are the discriminator and generator objective, respectively. $\mathcal{L}^\texttt{MMEL}$ is the mean square error (MSE) between the mel-spectrograms of the target and estimated speech signals with the number of mel filterbanks ranging from [64, 128, 256]. 6 STFT windows ([2, 4, 8, 16, 32, 64] ms) denoted by the STFT window index $k$ are used in our implementation.

\vspace{-0.25cm}
\section{Experiment Configurations}
\label{sec:config}


\subsection{Data}
Our experiments are conducted on the publicly available DNS-2023 challenge dataset. The training data contains $\sim$760 hour clean speech with speaker identity labels, $\sim$181 hour noise and $ \sim$60k room impulse response (RIR) samples. We use the DNS-2020 and DNS-2022 blind test set for evaluation.

\noindent\textbf{Preprocessing}: 
During the experiment, we found that the original single-speaker speech utterances may contain interfering speech from other speakers. The existence of interfering speech can confuse the PSE model in training and result in sub-optimal performance.
To address this issue, we divide the original utterances into 3-second segments and employ a pre-trained speaker embedding model to detect the interfering speech. 
This is achieved by computing the cosine similarity of speaker embeddings between each segment and the enrollment speech.
We removed segments with a similarity lower than 0.6 to clean up the data.

\noindent\textbf{Simulation}:
During training, the noisy speech signal is stimulated in an on-the-fly manner.
Specifically, we first sample one target speech and one noisy segment and then mix them with an SNR sampled from [-5, 20] dB. Note that we convolve 20\% of the speech segments with randomly selected RIR to simulate reverberation. 
For the PSE task, we additionally sample and mix an interfering speech segment with SIR ranging from [-5, 20] dB in simulation.
In PSE model training, 50\% of the training inputs are the mixture of target speech and noise segments, 30\% of the training inputs are the mixture of target, interfering speech and noise segments and 20\% of the training inputs are the mixture of target and interfering speech segments.
All the simulated training segments are set to be 6-second long in our implementation.

\begin{table*}[!ht]
    \centering
    \caption{Results on DNS-2020 \textit{no\_reverb} / \textit{with\_reverb} test sets. `-16k' denotes model trained using 16 kHz data. }
    \scalebox{0.95}{
    \begin{tabular}{c|c|c|cccc|cccccccc}
    \toprule
    \hline
    \multirow{2}{*}{ID} & \multirow{2}{*}{Model} & \multirow{2}{*}{Causal} & \multicolumn{4}{c|}{\textit{no\_reverb}}  & \multicolumn{4}{c}{\textit{with\_reverb}}              \\
    & &                               & PESQ-WB   & PESQ-NB        & SI-SNR    & STOI    & PESQ-WB       & PESQ-NB & SI-SNR  & STOI \\
    \hline
    1& Noisy               & $-$       & 1.58      & 2.45 & 9.1     & 91.5     & 1.82    &2.75  & 9.0     & 86.6           \\
    2& SN-NET \cite{zheng2021interactive}              & \xmark  & 3.39      & $-$  & 19.5     & $-$      & $-$     & $-$  & $-$     & $-$             \\ 
    3& DPT-FSNET \cite{dptfsnet}          & \xmark  & 3.26      & $-$  & 20.4    & 97.7     & 3.53    & $-$  &18.1    & 95.2           \\
    4& TridentSE \cite{yin2022tridentse}          & \xmark  & 3.44      & $-$  & $-$     & 97.9     & 3.50    & $-$ & $-$     & 95.2            \\
    5& MTFAA-NET \cite{zhang2022multi}          & \xmark  & \bf{3.52}      & 3.76 & $-$    & $-$        & $-$     & $-$ & $-$     & $-$              \\
    \hline
    6& BSRNN-16k             & \xmark     &3.45   &3.87   &21.1   &98.3   & \bf{3.72}  & \bf{4.03}  &19.1  &\bf{96.3} \\
    7& BSRNN                 & \xmark     &3.32   &3.79   &21.1   &98.0   & 3.54  &3.94  &18.5  &95.6  \\
    8& BSRNN-S               & \xmark     &3.42   &3.85   &21.3   &98.3   & 3.61  & 3.95 & 19.0 & 95.8\\
    9& BSRNN-S + MGD         & \xmark     &3.50   &3.85   &21.4   &98.4   & 3.63  &3.95  &19.0  &96.1 \\
    10& BSRNN-S + MRSD         & \xmark    &\bf{3.53}   &\bf{3.89}   &\bf{21.4}   &\bf{98.4}   & 3.68  &3.98  &\bf{19.2}  &\bf{96.3} \\
    \hline
    \hline
    11& CTS-Net \cite{li2021two}            & \cmark  &2.94       & 3.42 &18.0    & 96.7     &3.02     &3.47  &15.6   &92.7            \\
    12& GaGNet  \cite{li2022glance}              & \cmark  &3.17       & 3.56 &18.9    & 97.1     &3.18     &3.57  &16.6   &93.2            \\
    13& MDNet  \cite{li2022mdnet}             & \cmark  &3.18       & 3.56 & 19.2   & 97.2     &3.24     &3.59  &16.9   &93.6             \\
    14& FRCRN \cite{zhao2022frcrn}              & \cmark  &3.23       &3.60  & 19.8    & 97.7    & $-$     & $-$ & $-$     & $-$\\
    15& MTFAA-NET \cite{zhang2022multi}          & \cmark      & \bf{3.32}      & 3.63 & $-$     & $-$       & $-$     & $-$ & $-$     & $-$               \\
    \hline
    16& BSRNN-16k          & \cmark     &3.23       &3.73       &19.9       &97.7          &\bf{3.37}       &\bf{3.82}       &17.6 &\bf{94.8}   \\
    17& BSRNN              & \cmark     & 3.14	    &3.63	    &19.3       &  97.4        & 3.13      & 3.64      &16.8 & 93.9      \\

    18& BSRNN-S            & \cmark      & 3.26       &3.73       &20.0       &  97.6        & 3.30      & 3.77      & 17.5 & 94.5      \\
    19& BSRNN-S + MGD       & \cmark      & 3.27       &3.73       &20.1       &  97.7        & 3.33      & 3.78      & 17.4 & 94.5      \\
    
    20& BSRNN-S + MRSD      & \cmark     &\bf{3.32}   &\bf{3.77}  &\bf{20.5}  & \bf{97.8}    &\bf{3.37}  &3.80  &\bf{17.9} &94.7  \\
    \hline
    
    \bottomrule
    \end{tabular}
    }
    \label{tab:dns}
\end{table*}


\begin{table}[!ht]
    \centering
    \caption{PDNSMOS P.835 scores on the DNS-2022 blind test set, where SIG means speech quality, BAK means background noise quality, and OVRL means overall quality.}
    \scalebox{0.75}{
    \begin{tabular}{c|c|ccc|ccccccccccccccccccccccc}
    \toprule
    \hline
    \multirow{2}{*}{ID} &\multirow{2}{*}{Method} &\multicolumn{3}{c}{With Interference} & \multicolumn{3}{c}{Without Interference}\\
    &                &SIG   &BAK   &OVRL  &SIG    &BAK  &OVRL \\
    \hline
    1& Noisy           &3.97 &1.82 &2.23 &4.22 &2.30 &2.74 \\
    2& NSNet2\cite{braun2020data}          &3.47 &2.71 &2.44 &3.73 &4.18 &3.42 \\
    3& DeepFilterNet2\cite{schroter2022deepfilternet2}  &3.56 &2.94 &2.68 &4.09 &4.49 &3.85 \\
    4& TEA-PSE\cite{ju2022tea}         &3.69 &3.64 &3.16 &4.16 &4.44 &3.89 \\
    5& TEA-PSE2\cite{ju2023tea}        &3.86 &3.83 &\bf{3.36} &4.23 &\bf{4.50} &3.98 \\
    \hline
    \hline
    6& BSRNN-S         &3.76 &3.49  &3.05 &4.19 &4.51 &3.95\\
    7& pBSRNN-S       &3.66 &\textbf{3.91}	   &3.21     &4.17 &\bf{4.50} &3.94\\
    8& + MRSD          &\textbf{3.90} &3.81	 &\textbf{3.37} &\textbf{4.30} &\bf{4.50} &\textbf{4.04} \\
    \hline
    \bottomrule
    \end{tabular}
    }
    \label{tab:pdns2022}
\end{table}

\begin{table}[!ht]
    \setlength\tabcolsep{3pt}
    \centering
    \caption{Results of subjective listening, objective word accuracy and final scores on the blind test set of DNS-2023.}
    \scalebox{0.8}{
    \begin{tabular}{c|ccc|ccccccccccccccccccccccc}
    \toprule
    \hline
    \multirow{2}{*}{Method} &\multicolumn{3}{c}{Headset} & \multicolumn{3}{c}{Speakerphone}\\
                    &OVRL   &WACC   &SCORE  &OVRL   &WACC   &SCORE \\
    \hline
    Noisy           &1.22 &0.843 &0.449 &1.24 &0.857 &0.459 \\
    Baseline        &2.34 &0.687 &0.511 &2.38 &0.727 &0.537 \\
    pBSRNN-S + MRSD  &2.65 &0.724 &0.568 &2.66 &0.724  &0.570 \\
    \hline
    \bottomrule
    \end{tabular}
    }
    \label{tab:pdns2023}
\end{table}

\subsection{Model}

\noindent\textbf{Band-spilt Scheme}: 
We split the spectrogram into 33 subbands, including twenty 200 Hz bandwidth subbands for low frequency followed by six 500 Hz subbands and seven 2 kHz subbands. 

\noindent\textbf{Hyperparameters}: 
We use the Hanning analysis window for STFT and set the window length and shift to 20 ms and 10 ms for the 48 kHz model, and 32 ms and 8 ms for the 16 kHz model.   
We set the feature dimension to $N=96$ and $N=128$ for the 48 kHz and 16 kHz models, respectively. 
A six-layer band and sequence modeling module with 192 dimensional LSTM is used in our system.
The estimation module uses the 384-dimensional MLP with \textit{Tanh} activation function and a gated linear unit (GLU) \cite{dauphin2017language} output layer.
We use layer normalization \cite{ba2016layer} for offline configuration and batch normalization \cite{ioffe2015batch} for online configuration. 
The online pBSRNN model has a computational complexity of approximately 14.7G multiply-accumulate operations (MACs) per second. Using the ONNX runtime implementation on an Intel i5 2.50GHz CPU, the online PSE model achieves 0.42 real-time factor(RTF).

\noindent\textbf{Speaker Embedding Model}: The pretrained speaker embedding model \cite{chen2022sjtu, wang2023wespeaker}\footnote{\url{https://github.com/wenet-e2e/wespeaker/blob/master/docs/pretrained.md}} is used in our PSE system.

\subsection{Training Pipeline}
To perform non-personalized speech enhancement, we train the BSRNN model (without the speaker enrollment module) from scratch using the MR loss with an initial learning rate of $1e^{-3}$ for 400k iterations.
Then, we finetune the model with the MGD and MRSD training objectives for another 100k iterations.
The training of PSE model is similar to the non-personalized model, and the only difference is that the PSE model uses the parameters of the pretrained non-personalized model as initialization.

All the models are trained on 8 Nvidia P40 GPUs using Adam optimizer with 0.98 learning rate decay for every 20k updates. The model training will be stopped if the best validation result is not achieved in 20k consecutive iterations.

\vspace{-0.25cm}
\section{Experimental Results}
\label{sec:result}
\subsection{Non-personalized speech-enhancement}
Table~\ref{tab:dns} compares the performance between the proposed non-personalized SE BSRNN models and several top-ranking benchmark models on the non-blind test of DNS-2020 challenge.
It can be observed that: 
1) Compared with models trained using wide-band (16 kHz) data, the performance of models trained on full-band data with bi-directional band-level modeling degrades significantly (ID 6 v.s. 7, ID 16 v.s. 17), which supports the claim that the high-frequency distortion can degrade the full-band enhancement performance;
2) Using bi-directional and uni-directional band-level modeling on low-frequency and high-frequency subbands separately can improve the performance of full-band SE models (ID 7 v.s. 8, ID 17 v.s. 18);
3) The use of MGD and MRSD training objectives can further improve the performance of the SE models. Models trained using MRSD achieves the best SE performance; 
4) The best BSRNN model outperforms previous SE benchmark systems on both offline (ID 5 v.s. 10) and online (ID 15 v.s. 20) scenarios and achieves the SOTA overall performance.


\subsection{Personalized speech-enhancement}
In Table~\ref{tab:pdns2022}, we present the PDNSMOS ablation results of the proposed system on the DNS-2022 blind test set. 
It can be observed that: 
1) The non-personalized and personalized BSRNN models show comparable performance when there is no interfering speech, while the personalized model outperforms the non-personalized model for inputs without interference speech;
2) The use of MDR discriminator loss will degrade the PDNSMOS score, one possible explanation is that the MGD loss is particularly designed for the PESQ-WB metric, which can be inconsistent with PDNSMOS; 
3) After finetuning with the MRSD discriminator loss, both models can achieve better OVRL scores on the PDNSMOS metric; 
4) The proposed model can achieve comparable or better results against previous SOTA PSE systems (ID 5 v.s. 10).  

Table~\ref{tab:pdns2023} shows the results of the proposed pBSRNN in the DNS-2023 challenge. 
We found that:
1) The proposed pBSRNN system obtains better performance against the baseline systems from the challenge organizer.
2) While the proposed pBSRNN system can significantly improve the subjective OVRL scores, its use can largely reduce speech recognition word accuracy (WACC). This implies that current PSE models may introduce unexpected artifacts and excessively suppress target speech when eliminating background noise and other interfering speech.


\vspace{-0.25cm}
\section{Conclusion \& Limitation}
\label{sec:conclusion}
This paper presented the design of the BSRNN-based speech enhancement and personalized speech enhancement models and the corresponding training pipeline for full-band speech signals. With the modified band-level modeling strategy and the use of MGD and MRSD training objectives, the proposed systems achieved competitive performance on various SE and PSE benchmark datasets.
However, as discussed in Section~\ref{sec:conclusion}, the use of the SE model will severely degrade the speech recognition performance, indicating that current SE and PSE models still suffer from over-suppression and model distortion.
 In addition, we also find that when the timbres of the target speaker and the interfering speaker are similar, the model may fail to remove the interfering speech. 
Furthermore, if a channel mismatch occurs between the enrollment speech and the target speech, the model might mistakenly interpret them as distinct speakers.


\newpage
\bibliographystyle{IEEEtran}
\bibliography{refs}
  \end{document}